# Preserved Edge Convolutional Neural Network for Sensitivity Enhancement of Deuterium Metabolic Imaging (DMI)

Siyuan Dong[1], Henk M. De Feyter[2], Monique A. Thomas[2], Robin A. de Graaf[2,3], James S. Duncan[1,2,3]

[1]Department of Electrical Engineering, Yale University, New Haven, Connecticut
[2]Department of Radiology and Biomedical Imaging, Magnetic Resonance Research Center, Yale University School of Medicine, New Haven, Connecticut
[3]Department of Biomedical Engineering, Yale University, New Haven, Connecticut

**Running head**

Preserved Edge CNN for Deuterium Metabolic Imaging

**Correspondence**

Siyuan Dong

The Anlyan Center N309, 300 Cedar Street, New Haven, CT 06520-8042

Email: s.dong@yale.edu

**Word Count**

4999

## Abstract

**Purpose:** Common to most MRSI techniques, the spatial resolution and the minimal scan duration of Deuterium Metabolic Imaging (DMI) are limited by the achievable SNR. This work presents a deep learning method for sensitivity enhancement of DMI.

**Methods:** A convolutional neural network (CNN) was designed to estimate the $^2$H-labeled metabolite concentrations from low SNR and distorted DMI FIDs. The CNN was trained with synthetic data that represent a range of SNR levels typically encountered in vivo. The estimation precision was further improved by fine-tuning the CNN with MRI-based edge-preserving regularization for each DMI dataset. The proposed processing method, PReserved Edge ConvolutIonal neural network for Sensitivity Enhanced DMI (PRECISE-DMI), was applied to simulation studies and in vivo experiments to evaluate the anticipated improvements in SNR and investigate the potential for inaccuracies.

**Results:** PRECISE-DMI visually improved the metabolic maps of low SNR datasets, and quantitatively provided higher precision than the standard Fourier reconstruction. Processing of DMI data acquired in rat brain tumor models resulted in more precise determination of $^2$H-labeled lactate and glutamate + glutamine levels, at increased spatial resolution (from >8 to 2 μL) or shortened scan time (from 32 to 4 min) compared to standard acquisitions. However, rigorous SD-bias analyses showed that overuse of the edge-preserving regularization can compromise the accuracy of the results.

**Conclusion:** PRECISE-DMI allows a flexible trade-off between enhancing the sensitivity of DMI and minimizing the inaccuracies. With typical settings, the DMI sensitivity can be improved by 3-fold while retaining the capability to detect local signal variations.

# Introduction

Deuterium Metabolic Imaging (DMI) is a novel approach that relies on $^2$H MRSI in combination with administration of a $^2$H-labeled substrate (1). Because DMI is a relatively simple and therefore robust technique that provides information on active metabolism, the method has great potential to become a clinically used technique. In parallel, DMI has shown to be a practical preclinical tool to map metabolism in various organs of different animal models (1,2,3). As is the case for any medical imaging method, DMI's spatial resolution is ultimately limited by the achievable SNR. One strategy to increase the SNR is to apply DMI at ultra-high fields, because of the favorable magnetic field dependence of $^2$H NMR (4). However, an approach that is independent of magnetic field strength would benefit both clinical and preclinical DMI applications.

Recently deep learning has achieved significant success in medical imaging (5,6,7). Unlike traditional algorithms that use predefined rules to compute the output from the input, machine learning algorithms can learn the rules from a large number of training inputs and outputs. Because these rules, or mappings, are often too complicated to be described by formula, data-driven machine learning algorithms are preferred for solving such complex problems. Deep learning, a specific type of machine learning, learns the complex functional mappings by using neural networks (NN). NN have previously been applied to artifact removal (8,9) and quantification (10,11) in $^1$H-MRSI. Additionally, some preliminary works showed that NN have the potential to enhance sensitivity of DMI (12) and dynamic DMI (13).

Here we developed a deep learning method for sensitivity enhancement of DMI. Specifically, a deep convolutional neural network (CNN) was designed to estimate the concentration of $^2$H-labeled metabolites from low SNR and distorted DMI FIDs. Because there are not enough existing data for this new metabolic imaging technique, synthetic training data were generated. Additionally, we incorporated MRI-based edge-preserving regularization (14) into the processing, which uses the anatomical information of MRI as a spatial prior to further improve the metabolic mapping. The proposed method, which we named PReserved Edge ConvolutIonal neural network for Sensitivity Enhanced DMI (PRECISE-DMI), and its effects on precision and accuracy of mapping $^2$H-labeled metabolite levels were evaluated using simulation studies, and finally PRECISE-DMI was applied to in vivo data of glucose metabolism acquired in rat brain tumor models.

# Methods

## Convolutional neural network

The goal of using a NN is to best approximate a function $f$ that maps the input **x** to the desired output **y**, i.e. $\mathbf{y} = f(\mathbf{x}|\boldsymbol{\theta})$, where $\boldsymbol{\theta}$ represents the parameters in the NN. Unlike traditional algorithms that use predefined model-based rules to approximate $f$, the deep learning training process enables the NN to learn the function $f$ from a large set of training data. A sizable number of trainable parameters in the NN must be learned to approximate the function. During training, an optimal set of parameters are found using gradient-based techniques, computed on a predefined loss function using a strategy known as backpropagation. The trained NN can then be used to map any new input to the output. There are mainly two issues to be solved in a deep learning task, namely the designs of 1) a proper NN architecture that could approximate function $f$ in a meaningful and efficient way, and 2) a training methodology including training data, loss function and hyperparameters. The second issue is discussed in later sections.

To address the first issue, the nature of the data must be observed. The goal is to estimate the $^2$H-labeled metabolite concentrations from noisy FID inputs. The NN architecture should be devised in a way that it takes complex FID time samples as an input and puts out a vector of metabolite amplitudes. A CNN contains trainable convolution kernels to extract features from the input and has demonstrated excellent performance, i.e. fast convergence, for this problem. A large number of convolution kernels were stacked in layers to model the functional mapping from the input to the output. Because convolution is a linear operation, a nonlinearity PReLU (parametric rectified linear unit) (15) was introduced after each convolution layer to model nonlinear mappings. Additionally, the max pooling layers were used for down-sampling. The outputs of different layers represent different feature levels and are called "feature maps". The final feature map was flattened and fed through two fully connected (FC) layers to produce the final output. The FC layers contain trainable weights and bias to perform multiplication and addition. The proposed CNN architecture is shown in Figure 1.

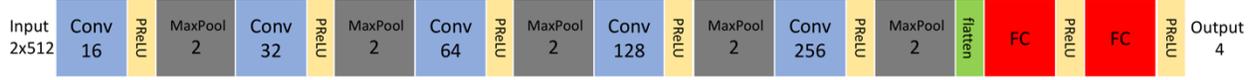

*Figure 1: The proposed CNN architecture. The input matrix contains real and imaginary parts of FID time samples (represented here as 2×512 matrix). Conv, PReLU, MaxPool, FC represents convolution (with the number of kernels underneath), parametric rectified linear unit, max pooling (with the window size underneath) and fully connected layers respectively. The convolution layers use kernel size of 3. The output size depends on the number of metabolites. In this work, the output size is 4 (3 metabolites plus water).*

## Training data generation and single-voxel estimation

Due to a lack of abundant in vivo DMI data, training data for the CNN were generated synthetically. The parametric model of an ideal DMI FID is defined as

$$\mathrm{x}(t) = \sum_{m=1}^{M} A_m \exp(2\pi i f_m t) \exp\left(-\frac{t}{T_m}\right) \quad [1]$$

where $M$, $A$, $f$, $T$ and $t$ represent the number of metabolites, amplitude, frequency, T$_2$* relaxation time constant and discrete time samples, respectively. However, an in vivo acquired FID also involves several imperfections: Gaussian white noise, phase distortion and B$_0$ inhomogeneity effects. To take these factors into account, a realistic DMI FID can be modeled with

$$\mathrm{x}'(t) = \sum_{m=1}^{M} A_m \exp(2\pi i (f_m + \delta f)t) \exp(i\varphi) \exp\left(-\frac{t}{T_m + \delta T}\right) + N(t), \ N(t) \sim \mathcal{N}(0, \sigma) \quad [2]$$

where $\delta T$ and $\delta f$ model line-broadening and frequency shift caused by the B$_0$ inhomogeneity effects. $\varphi$ models the random phase distortion and $N(t)$ is Gaussian noise with SD $\sigma$. The objective for the CNN is to learn an inverse mapping from $\mathrm{x}'(t)$ to metabolite amplitudes $A_1, A_2, \ldots, A_m$.

To generate training data, each of the training outputs $A_1, A_2, \ldots, A_m$ was sampled from a uniform distribution over a range $[0, A_{max}]$ where $A_{max}$ is 2 times larger than the maximum amplitude expected in experimental DMI data. $f_1, f_2, \ldots, f_m$ were fixed and based on the known $^2$H chemical shifts of the metabolites. $T_1, T_2, \ldots, T_m$ were fixed and based on previously published values (1). The additional line broadening parameter $\delta T$ and frequency shifts $\delta f$ are study-dependent and were sampled from uniform distributions over the range $[\delta T_{min}, \delta T_{max}]$ and $[\delta f_{min}, \delta f_{max}]$ respectively, where the boundaries were set based on realistic values observed in vivo. The phase distortion parameter $\varphi$ was sampled from $[0, 2\pi]$. Finally, training input FIDs were generated from the sampled parameters using Equation 2.

The CNN was trained with the training input-output pairs using the mean squared error loss function

$$L_1 = \frac{1}{N} \sum_{i=1}^{N} \|\mathcal{C}(\mathbf{X}_i) - \mathbf{A}_i\|_2^2 \qquad [3]$$

where $N$ is the number of training pairs, $\mathcal{C}(\mathbf{X})$ is a vector containing the metabolite amplitudes estimated by the CNN from noisy and distorted (phase-uncorrected, line-broadened and frequency-shifted) FID time samples contained in $\mathbf{X}$, and $\mathbf{A}$ is a vector containing ground truth amplitudes. The training data generation and training process are summarized in Figure 2A. The trained CNN was then used for estimating the metabolite amplitudes for any unseen experimental DMI at each individual spatial location, for which we use the term single-voxel estimation (SVE).

## Edge-preserving regularization

To further improve the estimation precision for a specific DMI dataset, we incorporated corresponding spatial information from high resolution MRI via the edge-preserving regularization. Here we used a technique called fine-tuning, which takes a NN pretrained with a global dataset and applies it to a subtask. During fine-tuning, the parameters in the convolution layers are normally fixed and those in the FC layers are retrained with the subtask data, which avoids retraining the entire NN and hence accelerates the optimization process during learning. In this work fine-tuning was applied to incorporate the spatial information. Specifically, a new data-specific CNN (same structure as the SVE CNN) was initialized with the parameters in the trained SVE CNN, and its FC layers were fine-tuned to minimize the loss function

$$L_2 = \sum_{n_1=1}^{V} \left\| \mathcal{C}_2(\mathbf{Y}_{n_1}) - \mathcal{C}(\mathbf{Y}_{n_1}) \right\|_2^2 + \lambda \sum_{n_2 \in \Omega_{n_1}} \omega_{n_1 n_2} \left\| \mathcal{C}_2(\mathbf{Y}_{n_1}) - \mathcal{C}_2(\mathbf{Y}_{n_2}) \right\|_2^2 \qquad [4]$$

where $V$ is the number of voxels in the DMI, $\lambda$ is the regularization parameter, $\Omega_{n_1}$ represents the neighboring voxels of a voxel $n_1$, $\omega_{n_1 n_2}$ is the spatial coefficient for a voxel pair $(n_1, n_2)$, and **Y** contains time samples of a noisy and distorted FID. $\mathcal{C}()$ and $\mathcal{C}_2()$ are vectors containing the metabolite amplitudes estimated by the trained SVE CNN and the fine-tuned data-specific CNN, respectively. The first term forces the output of the data-specific CNN to be faithful to the output of the SVE CNN, and the second term is the edge-preserving regularization that imposes anatomical priors from MRI. The MRI-based edge-preserving regularization (14) assumes that there is some correlation between the MRI intensity (differences) and the underlying MRSI metabolite concentrations (differences). The practical implementation establishes the MRI intensity difference between two neighboring voxels ($n_1$ and $n_2$). A low MRI signal difference, being synonymous with the absence of an edge, is used to impose a greater level of smoothness on the underlying metabolite levels. When the signal difference is high (i.e. in the presence of an edge in the MR image), the smoothness is not enforced in order to preserve the underlying metabolite heterogeneity. The amount of smoothness is controlled via the spatial coefficients $\omega_{n_1 n_2}$:

$$\omega_{n_1 n_2} = \min\left( \frac{1}{\left( r_{n_1} - r_{n_2} \right)^2}, \omega_{max} \right) \qquad [5]$$

where $r_{n_1}$ and $r_{n_2}$ correspond to the preprocessed MRI gray levels at voxel $n_1$ and $n_2$ respectively, and $\omega_{max}$ serves to avoid over-smoothing. In this way, the estimation precision was further improved (compared to performing SVE alone), especially for the voxels having large spatial coefficients with their neighbors. Figure 2B shows the details of the regularization process.

In summary, the trained SVE CNN was used for all DMI that have similar metabolic content, and the data-specific CNN was fine-tuned with the loss function in Equation 4 to process each individual experimental DMI. Although the MRI-based edge-preserving regularization can further improve the estimation precision, one concern is that it might lead to inaccurate results. When the regularization is weighted less, PRECISE-DMI result will be faithful to the original DMI, but the level of precision improvement will be limited. When the regularization is more heavily weighted, the precision will be higher, but the result can differ from the underlying DMI. In an extreme scenario, PRECISE-DMI result will instead look very similar to the MRI, where there is no metabolic information at all.

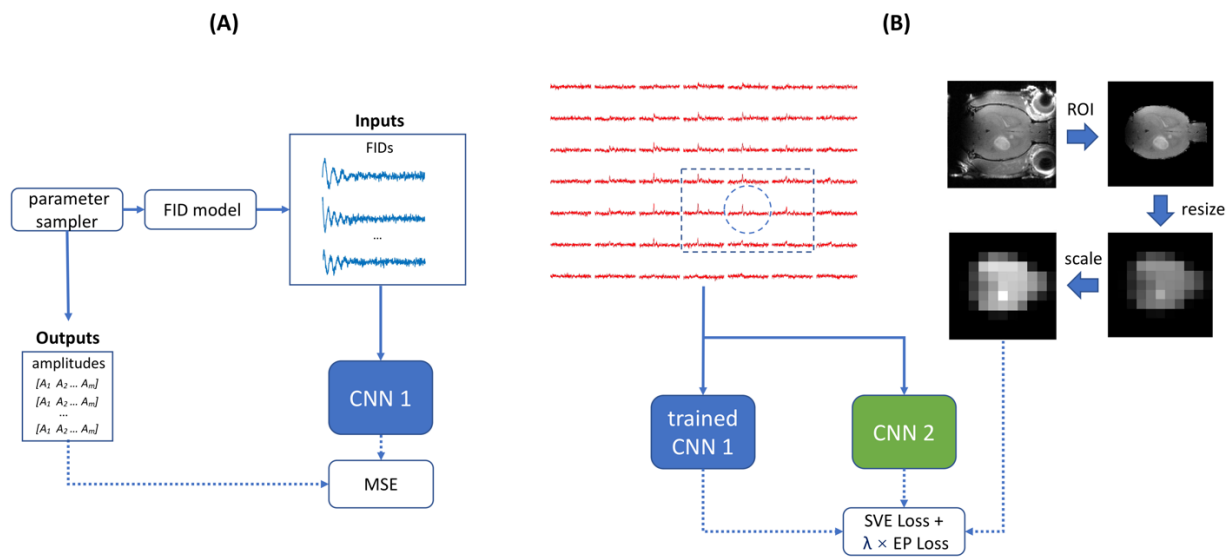

*Figure 2: Overview of the proposed processing method of PRECISE-DMI. (A) Training data generation and training of the SVE CNN (CNN 1). The "FID model" is described in Equation 2, and the mean squared error (MSE) loss function is described in Equation 3. (B) Fine-tuning with the edge-preserving (EP) regularization. Firstly, to be used for computing spatial coefficients described in Equation 5, the high-resolution MRI was preprocessed by 1) applying a binary mask that keeps only the ROI (voxels inside the brain), 2) down-sampling from its original matrix size to the DMI matrix size, and 3) rescaling the gray levels to the range [0,1]. After that, the parameters in the FC layers of the data-specific CNN (CNN 2) were fine-tuned with the loss function (SVE Loss + $\lambda \times$ EP Loss) described in Equation 4. To process each individual voxel (dashed circle), its local 3×3 (or 3×3×3 for 3D) neighboring voxels (dashed rectangle) were grouped into a batch and sent to the CNNs. The loss function was minimized within each batch during the fine-tuning process.*

## Data acquisition
### I. Simulation studies

The FIDs used for simulation studies contain four resonances modelling four compounds that are typically observed in DMI studies using [6,6-$^2$H$_2$]-glucose as a metabolic substrate. These compounds include water, glucose (Glc), glutamate and glutamine (Glx), and lactate (Lac) at chemical shifts 4.7, 3.7, 2.4, 1.3 ppm respectively. The simulation phantom (Figure 3) contains three ring-like compartments and each demonstrates a distinct metabolite profile. In addition, to evaluate the effect of the edge-preserving regularization, we introduced inconsistencies between DMI and MRI at four small tumor-like compartments denoted as A, B, C and D. In compartment A, there is no tumor in neither MRI nor DMI; in B, there is no tumor in MRI, but DMI shows a metabolic abnormality; in C, a tumor is visible in both MRI and DMI; in D, the tumor is visible in MRI, but DMI shows normal metabolism. The compartments with metabolic abnormality demonstrate lower Glx and higher Lac concentrations than other compartments, consistent with observations in high grade brain tumors (1). With MRI as the diagnostic readout and DMI as the ground truth, compartments A, B, C and D are defined as true negative, false negative, true positive and false positive respectively.

In a separate set of simulations, $B_0$ and $B_1$ inhomogeneities were also incorporated into the simulated DMI. The incorporation of $B_0$ inhomogeneity brought frequency shift and line-broadening effects to the simulated FIDs, and the incorporation of $B_1$ inhomogeneity introduced signal amplitude variations across the phantom. The designs of $B_0$ (Figure 6A) and $B_1$ maps were all based on realistic ranges and gradients that have been encountered through experimentation.

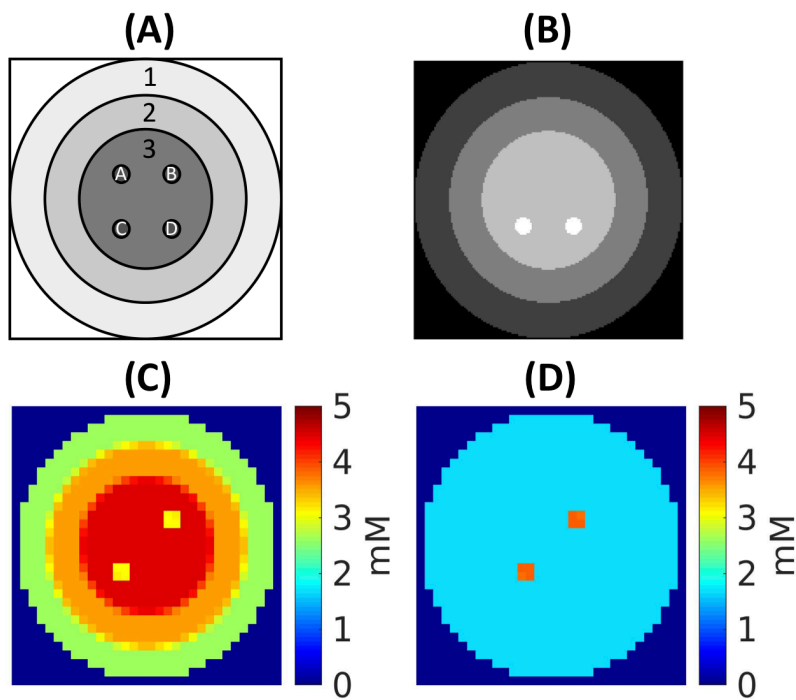

*Figure 3: Design of the simulation phantom. (A) Compartment design; (B) MRI; (C) DMI-derived Glx concentration; (D) DMI-derived Lac concentration. The simulated DMI dataset was constructed with the following parameters: 160×160 mm$^2$ FOV, 32×32 matrix, 512 FID time samples and 2.5 kHz spectral width; the simulated MRI was constructed with the same FOV and with a matrix size of 160×160. In this paper, all metabolite signals (Glc, Glx and Lac) are expressed as a ratio to the water signal. The ratio is converted to concentration (in mM) by using the "natural abundance" $^2$H water concentration of 10 mM. Additionally, the SNR is defined as the peak height of water divided by the noise SD in the spectral domain, since the amplitude of water is homogeneous across the entire phantom while the amplitudes of the metabolites are not.*

**II. In vivo experiments**

Animals studies were performed on an 11.74T Magnex magnet (Magnex Scientific Ltd, Oxford, UK) interfaced to a Bruker Avance III HD spectrometer, running on Paravision 6 (Bruker Instruments, Billerica, MA, USA). The system is equipped with 9.0 cm diameter Magnex gradients capable of switching 440 mT/m in 180 μs and 0.5 kW RF amplifiers for $^1$H and $^2$H transmission.

RF transmission and reception for MRI and shimming during brain studies was performed with a quadrature 24 mm diameter surface coil tuned to the proton NMR frequency (499.8 MHz). Deuterium RF transmission and reception was achieved with a 20×15 mm$^2$ diameter surface coil tuned to 76.7 MHz.

Deuterium-labeled glucose ([6,6'-$^2$H$_2$]-glucose) was purchased from Cambridge Isotopes Laboratories, Inc, (Tewksbury, MA, USA), dissolved in water at a concentration of 1 M. Infusions resulted in 1.95 g/kg body weight of [6,6'-$^2$H$_2$]-glucose infused during a 120 min study. Infusions were performed via an intravenous line in the femoral vein or an intraperitoneal infusion line. All animal procedures were approved by the Yale University Institutional Animal Care and Use Committee.

Fischer 344 rats were used for DMI experiments after intracerebral implantation of RG2 cells, as described previously (1). For DMI experiments, rats were anesthetized with isoflurane using ~60% O$_2$ and ~40% N$_2$O delivered through a nose-cone. A heating pad was used to maintain body temperature at ~37 C, and respiration was monitored via a pneumatic pillow (SA Instruments, Stony Brook, NY, USA).

$^2$H MR signal acquisition was achieved with a pulse-acquire sequence extended with 3D phase-encoding gradients during the initial 0.6 ms following excitation. All data were acquired with a repetition time TR of 400 ms over a 5 kHz spectral width. With spherical k-space encoding a typical 11×11×11 matrix was acquired in circa 4 min per signal average. To test the sensitivity enhancement given by PRECISE-DMI, DMI datasets were acquired with 1, 2, 4 or 8 averages, and with voxel sizes ranging from 2 μL to 16 μL. In all cases, the corresponding Gd-enhanced, T$_1$-weighted, spin-echo MRI (TR/TE = 1000/6 ms) was acquired over the identical FOV with a 110×110×110 matrix. All DMI acquisitions were performed during metabolic steady state, typically between 90 and 120 min following the onset of glucose infusion.

## Evaluation

Apart from the visual quality improvement, the performance of PRECISE-DMI was evaluated for accuracy (bias) and precision (SD). In simulation studies, the ground truth was available so bias and SD were computed directly from repeated experiments with Gaussian noise added to the ground truth. Precision and accuracy can only be estimated indirectly for in vivo experiments, therefore the SVE result was used in place of the ground truth. The error estimation for in vivo experiments involved three steps: 1) perform SVE (λ = 0), 2) repetitively add noise to the FIDs reconstructed from the SVE result, and 3) carry out full PRECISE-DMI (λ > 0). Based on the unbiased nature of SVE in the normal SNR range, the bias imposed by the edge-preserving regularization was approximated by the absolute difference between the average PRECISE-DMI result and the SVE result.

## Training and implementation details

For all experiments, training FIDs have 4 peaks (water, Glc, Glx and Lac), with frequencies and relaxation time constants set to corresponding published values (1). Different levels of Gaussian noise were introduced to each training FID: the noise SD σ (Equation 2) was sampled from a uniform distribution, i.e. $\sigma \sim \mathcal{U}(0, \sigma_{max})$. $\sigma_{max}$ was chosen to effectively simulate poor SNR data. 80% of the training FIDs had added noise, while the other 20% were noiseless. The noiseless FIDs helped the NN to find more accurate inverse mapping from FIDs to metabolite amplitudes, and thus reduced an occasional <2% bias in metabolite amplitude estimation that was observed when all training FIDs had added noise.

Because the training FIDs were randomly generated during the training process, each training iteration relied on new and unused training data. Therefore, overfitting was not a problem and large numbers of training iterations were used. Empirically, batch size and number of iterations were determined to be 128 and 25000, respectively (equivalent to 3.2 million training data), after which the loss function flattened. Experiments were implemented in PyTorch on a NVIDIA's GeForce GTX 1080 GPU. Both training and fine-tuning were performed with Adam optimizer (16), and with initial learning rate of 0.001. The training took ~17 min and the computation time of the fine-tuning process depends on the matrix size of the experimental DMI data.

## Results

### Simulation studies

#### I. Analyses of single-voxel estimation

Figure 4 shows the performance of CNN-based SVE for Glx and Lac at three different SNR levels (18.6, 12.1 and 7.7). CNN-based SVE improved the compartmental resolution and extracted small tumors from low SNR DMI. The SD maps and the SD-shaded spectra show improved estimation precision compared to the standard Fourier reconstruction (peak integral in the spatiospectral domain of the raw data) at all three SNR levels. However, a small noise-induced bias was observed from the spectrum at the lowest SNR, especially for Lac. Note that a SNR of 7.7 for water corresponds to a SNR of <1 for Lac, since the water level is ten times higher than Lac.

To quantify the improvement in SNR, a Monte Carlo simulation (N = 1000) on a voxel in compartment 3 was carried out for 14 SNR levels over the range [5, 35]. Mean and SD of the percentage error are shown in Figure 5A. The graph illustrates a ~1.7-fold improvement in SNR achieved by CNN-based SVE, compared to Fourier reconstruction (for example, SD of the SVE result at SNR = 12.1 is roughly that of Fourier reconstruction at SNR = 20.5).

The performance was also compared with the Cramér-Rao lower bound (CRLB) and the commonly used model-based spectral fitting approach. CRLB is a lower bound on the SD of the estimated parameters from an unbiased estimator. The method for how to derive CRLB on parameter estimation for a summation of exponentially damped sinusoids in presence of Gaussian noise was described in previous literature (17). In our setup, there are four nonoverlapping peaks, and they have equal phases, fixed relative frequencies and fixed relative linewidth. Given these prior knowledge, the CRLB of amplitude estimation was derived as

$$\text{CRLB}_A = \sqrt{2} T^{-\frac{1}{2}} \sqrt{t_s} \sigma R \qquad [6]$$

where $T$, $t_s$, $\sigma$ and $R$ are the relaxation time constant, the sampling time interval, the time-domain noise SD and an amplitude-dependent term respectively. Here the Glx was investigated and $R = ((A_{water}^2 + A_{Glc}^2 + 2A_{Glx}^2 + A_{Lac}^2)/(A_{water}^2 + A_{Glc}^2 + A_{Glx}^2 + A_{Lac}^2))^{1/2}$. Figure 5B shows that SD of CNN-based SVE and of the spectral fitting were both very close to the CRLB, which means that the ultimate precision of unbiased estimation was almost achieved. For this particular data, CNN-based SVE took ~1 second, whereas the spectral fitting took ~80 seconds, using the same CPU for both computations.

The spectrum in Figure 6C shows that CNN-based SVE could deal with significant frequency shift and line-broadening effects caused by $B_0$ inhomogeneity. However, the amount of improvement in SD was affected by the magnitude of the $B_0$ gradient. The estimation precision was lower in voxels where the $B_0$ gradients were large (Figure 6B, marked with the red circle) because the SNR was lower due to the line-broadening effect. The performance under $B_1$ inhomogeneity was also investigated (data not shown). The $B_1$ inhomogeneity had no additional effect on the performance of CNN-based SVE, except that the estimation precision was reduced due to the lower signal sensitivity (lower SNR) in areas with lower $B_1$ values.

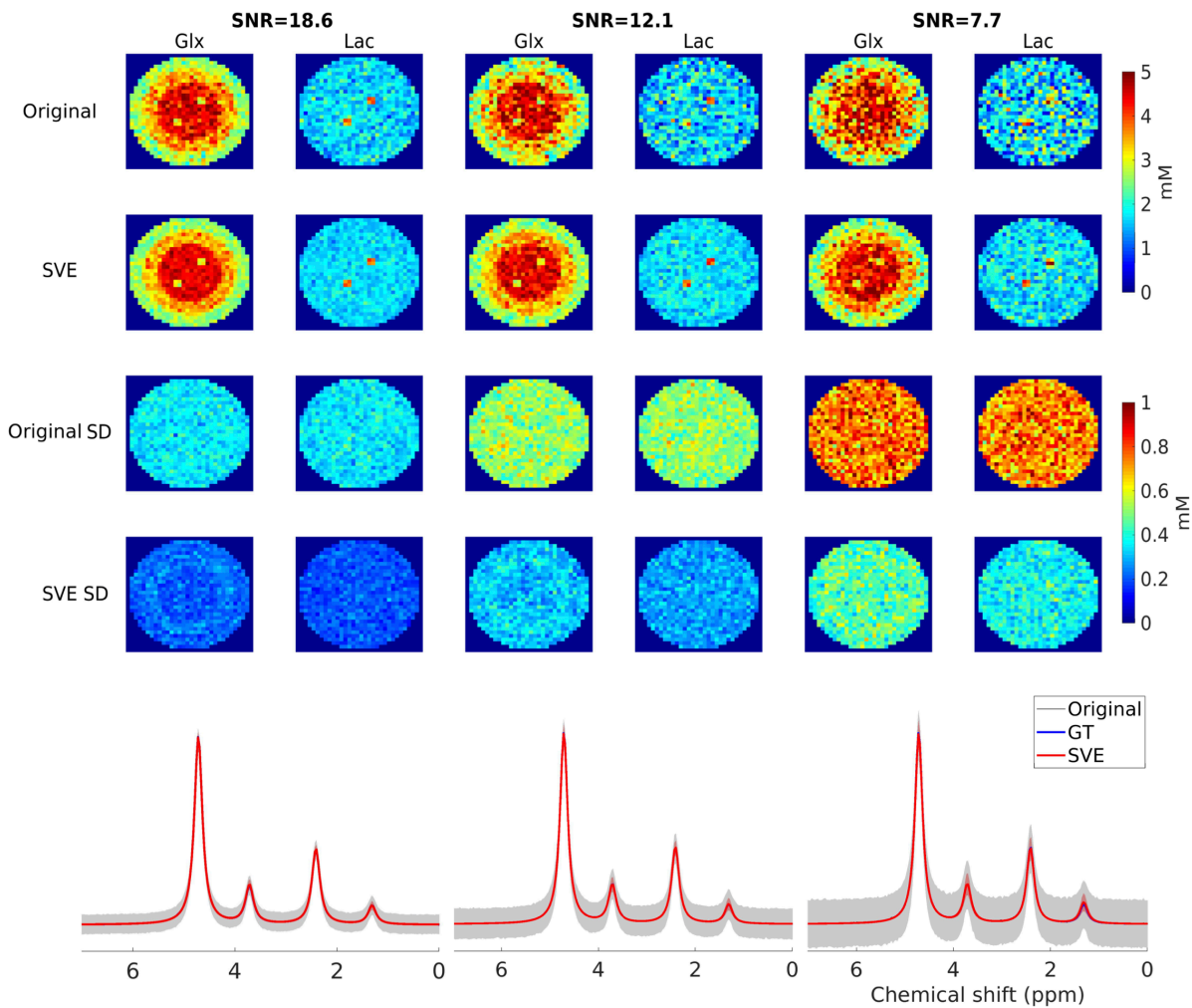

*Figure 4: CNN-based SVE results. From left to right are the results at three different SNR levels (18.6, 12.1 and 7.7), for Glx and Lac. First row: concentration maps after Fourier reconstruction (Original); Second: concentration maps after SVE; Third: SD maps after Fourier reconstruction (Original SD); Fourth: SD maps after SVE; Fifth: corresponding spectra (GT = ground truth) from a voxel in compartment 3. Spectral values are shown as mean $\pm$ SD (shaded area).*

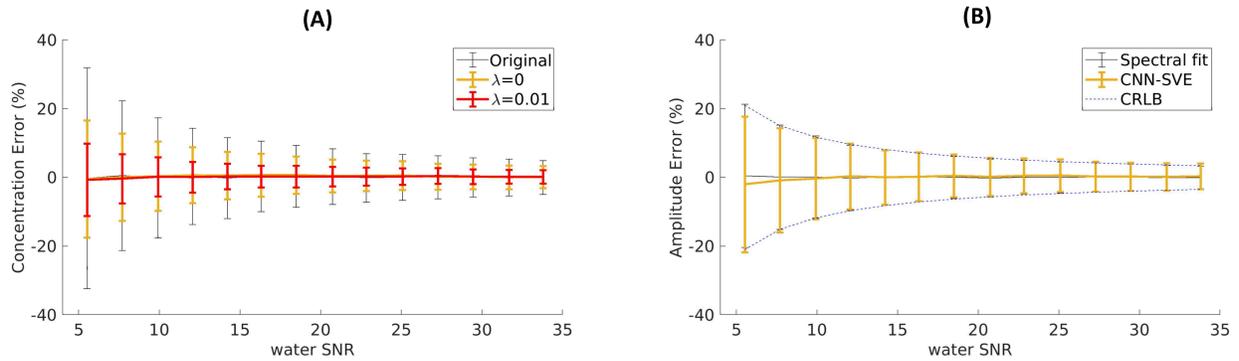

*Figure 5: Monte Carlo simulations. (A) Percentage error of Fourier reconstruction (Original), CNN-based SVE (λ = 0) and PRECISE-DMI (λ = 0.01); (B) Comparison of model-based spectral fitting, CNN-based SVE and CRLB. The errors are shown in mean ± SD. Note the different vertical axis in (B) compared to (A), because CRLB provides bounds on estimation precision of peak amplitudes, not of peak ratio. The investigated metabolite is Glx.*

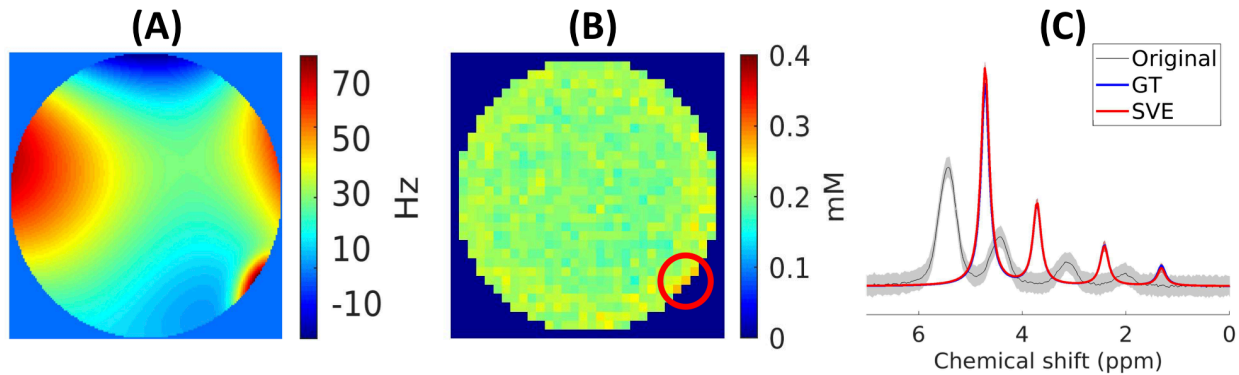

*Figure 6: Performance of CNN-based SVE when dealing with $B_0$ inhomogeneity. (A) The simulated $B_0$ map; (B) SD map after SVE, for Lac; (C) Spectra of Fourier reconstruction (Original), ground truth (GT) and SVE, from a voxel with strong frequency shift and line-broadening effects (inside the red circle). Spectral values are shown as mean ± SD (shaded area).*

## II. Analyses of edge-preserving regularization

Figure 5A shows that with a moderate level of spatial information incorporation (λ = 0.01), the SNR were further improved (~3-fold). This result was obtained in a spatially homogeneous location with a perfect match between MRI and DMI (compartment 3).

To understand the potential effects of the edge-preserving regularization, we evaluated the bias and SD trade-off over all spatial locations, and with three different values of the regularization parameter: $\lambda$ = 0.004, 0.01 and 0.04.

Figure 7 shows that imposing a heavily weighted regularization ($\lambda$ = 0.04) visually greatly improved the image quality. The metabolic maps given by PRECISE-DMI with $\lambda$ = 0.04 were less noisy and the compartment borders were clearer than those with $\lambda$ = 0.004 and 0.01. In addition, a more heavily weighted regularization (larger $\lambda$) gave higher precision (lower SD). Note that the SD at the edge voxels were higher than at voxels not near an edge because the regularization preserved voxels where the MRI indicated edges. On the other hand, the edge-preserving regularization increased the bias significantly in compartment B where there was an MRI-DMI mismatch, i.e. the false negative result. At this location, the pattern of high Lac and low Glx was less pronounced and appeared to be closer to the Lac and Glx levels of the surrounding compartment because there was no edge prior in the MRI at that location. When a large $\lambda$ (0.04) was used, the false negative tumor disappeared almost completely. Additionally, even the true positive tumor at compartment C started to have a significant bias because of the "over-smoothing" effect: the nonzero spatial coefficients ($\omega$ in Equation 2) around this tumor were amplified to an over-smoothing level. That also explains why there was unwanted bias for Glx in compartment 1 and 3.

The bottom two rows in Figure 7 show that the errors estimated by the proposed error estimation method for in vivo experiments resembled the real errors. Although the estimated bias maps were noisier than those based on the ground truth, the position and magnitude of bias were still informative.

To analyze the effects of tumor size (Figure 8), we carried out simulation experiments with modified tumor sizes of 1×1, 2×2 (original) and 4×4 voxels, using $\lambda$ = 0.01. When the tumor was large (4×4), the SD was higher for the edge voxels than for the central voxels of the true positive tumor because the regularization preserved edges. Similar to previous analyses on a 2×2 tumor, a 4×4 false negative tumor was still challenging, but the bias was smaller because there were more tumor voxels clustered together, which counteracted the smoothing. However, when the tumors were only visible in a single voxel (1×1), the false negative tumor was almost completely smoothened out by the edge-preserving regularization and the bias at that location was larger than that of 2×2 and 4×4 tumors.

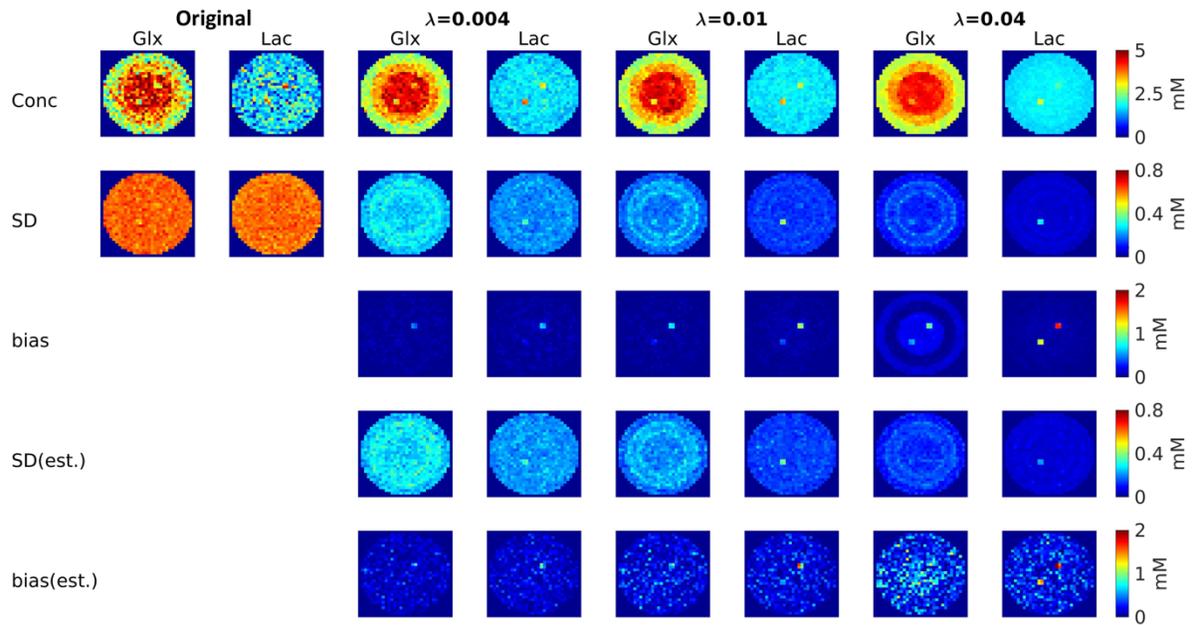

*Figure 7: Effects of the edge-preserving regularization under different values of regularization parameter. From left to right are results of Fourier reconstruction (Original) and PRECISE-DMI with λ = 0.004, 0.01 and 0.04, for Glx and Lac. From top to bottom are concentration (Conc), SD, bias, estimated SD and estimated bias maps.*

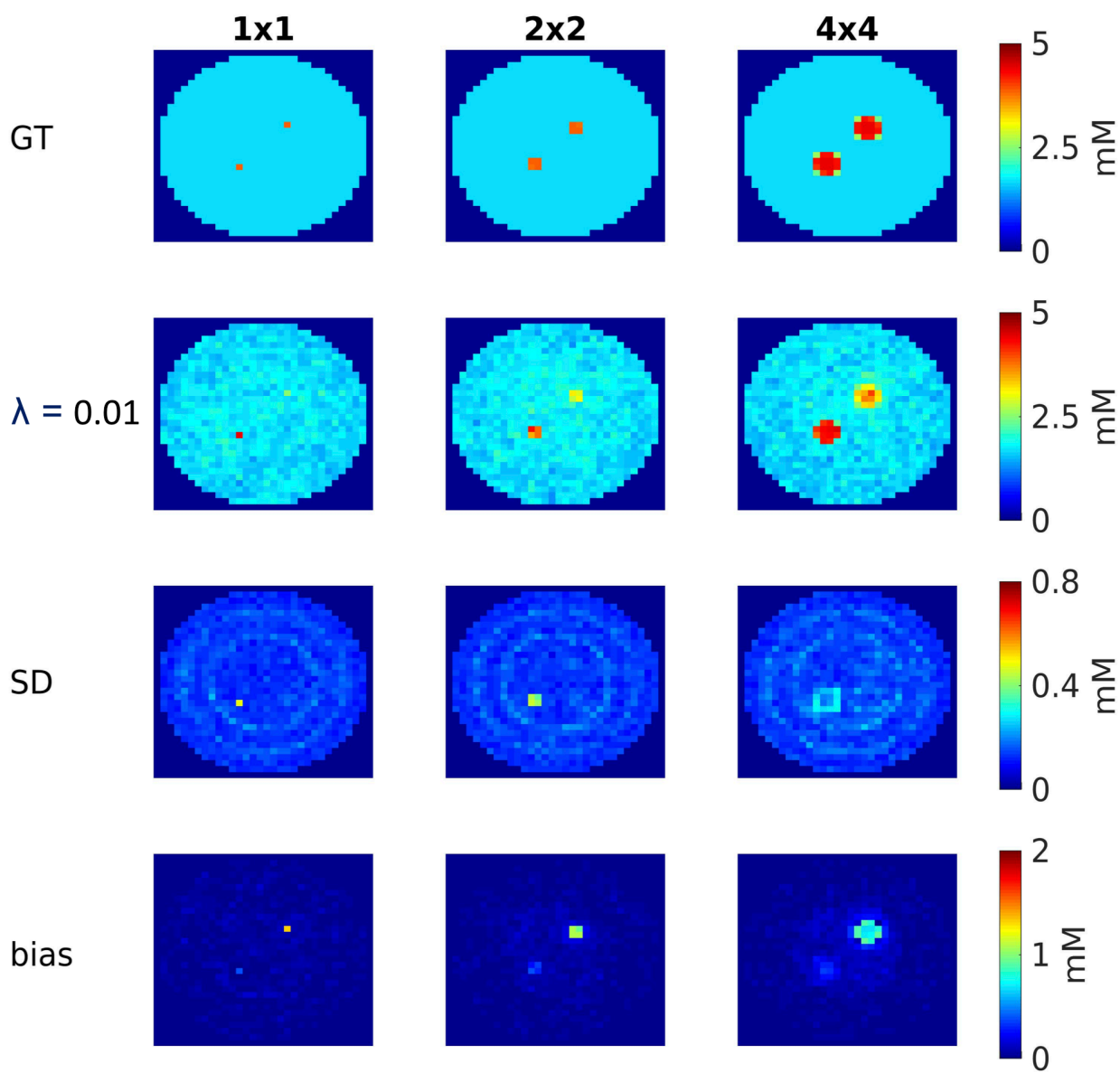

*Figure 8: Effects of tumor size. From left to right are results for tumor sizes of 1×1, 2×2 and 4×4 voxels. The first row shows ground truth (GT) concentration maps. The following rows show concentration, SD and bias maps after performing PRECISE-DMI with $\lambda$ = 0.01. The investigated metabolite is Lac.*

## In vivo experiments

For in vivo experiments, we first analyzed the potential of PRECISE-DMI on improving SNR for data acquired with relatively low SNR but high spatial resolution of 2 μL. Because the data was subject to significant frequency shift and line-broadening effects, it was hard to perform direct peak integral in the Fourier reconstructions. Therefore, the results of PRECISE-DMI were compared with those obtained from performing spectral fitting. The analyses were repeated with $\lambda$ = 0.01 and 0.04. As shown in Figure 9, the metabolic maps obtained from spectral fitting were still quite noisy, especially at the brain periphery and in the region of the olfactory bulb due to relatively lower $B_1$ field and stronger $B_0$ inhomogeneity in those areas (Figure 9D). The metabolic maps given by PRECISE-DMI with $\lambda$ = 0.01 had fewer noise-induced variations across the brain and indicated high Lac concentration only at the tumor position. As expected, $\lambda$ = 0.04 gave lower estimated SD, but the estimated bias at the tumor became larger because of the over-smoothing effect. To verify how PRECISE-DMI handles a false negative in vivo, the image contrast of the tumor from the contrast-enhanced MRI was manually removed (Figure 9B). In this false negative setting, PRECISE-DMI with $\lambda$ = 0.04 produced significant bias, which resulted in a very low Lac/Glx ratio at the tumor (Figure 9E). These observations are consistent with the simulation results in Figure 7.

We also demonstrated the potential of PRECISE-DMI on improving SNR for data acquired with relatively short scan time. Figure 10 shows the metabolic maps and the estimated SD maps of Lac for 4, 2 and 1 average (NA) acquisitions, representing scan times of 16, 8 and 4 min, respectively. The experiments were carried out with $\lambda$ = 0.01, which was decided to be a reasonable value in previous simulation and in vivo experiments (Figure 9). At the highest SNR (NA = 4) the Lac map obtained from performing spectral fitting showed the tumor at the correct position, but there were some suspicious areas at the brain periphery where the $B_1$ field was low and $B_0$ gradient was large. After PRECISE-DMI was applied, the map exhibited high Lac concentration only at the tumor position, and the remaining regions were quite homogeneous. As shorter scan times were used (NA = 2 and 1), the performance was deteriorating, but there were still significant improvements over spectral fitting. One can observe that after processed by PRECISE-DMI, data acquired with the shortest scan time (NA = 1) were more precise than the unprocessed data acquired with the longest scan time (NA = 4), which means the scan time can be reduced while still maintaining similar SNR.

For comparison, another edge-preserving denoising technique, anisotropic diffusion (18), was applied to the metabolic maps obtained from spectral fitting at NA = 1. Anisotropic diffusion encourages smoothing within homogeneous regions but not across edges, and edges are detected automatically based on image gradient. The edge-detection distinguishes edges from noise based on a gradient threshold (chosen experimentally), and higher threshold imposes more smoothing. With a threshold of 10% (of the range of voxel values), anisotropic diffusion failed to correct suspicious areas at the brain periphery (Figure 10D). While a larger threshold of 20% could correct some of the suspicious areas, the tumor contrast started to be reduced.

PRECISE-DMI requires short computation time: processing a 11×11×11 in vivo DMI data took ~31 seconds for the fine-tuning process to sufficiently converge.

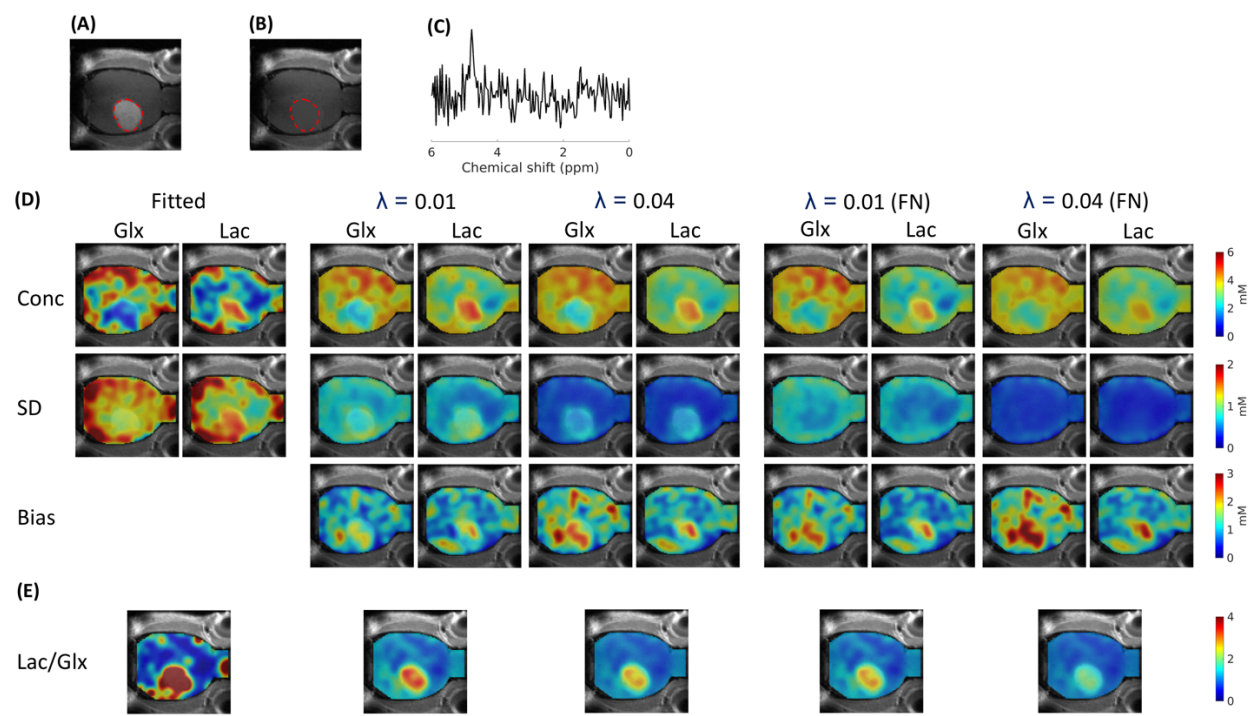

Figure 9: In vivo experiment results: Improving SNR for DMI data with high spatial resolution. The DMI data was acquired with a standard scan time of 32 min (NA = 8) but a high spatial resolution of 2 $\mu L$ (22×11×22 matrix over 22×22×22 $mm^3$ FOV). (A) MRI (tumor circled with the red dashed line); (B) MRI with image contrast of tumor removed; (C) $^2H$ MR spectrum from a tumor voxel; (D) From left to right are the results obtained from spectral fitting (Fitted) and PRECISE-DMI with $\lambda$ = 0.01 and $\lambda$ = 0.04, with and without the false negative (FN) setting. From top to bottom are concentration (Conc), estimated SD and estimated bias maps. (E) The ratio of Lac/Glx for corresponding experiments in (D).

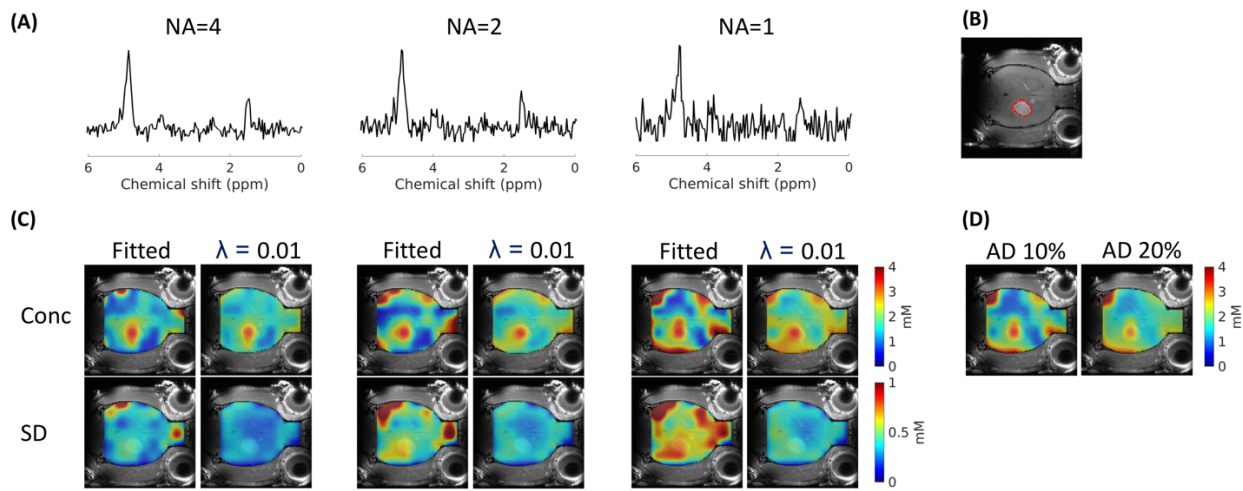

*Figure 10: In vivo experiment results: Improving SNR for DMI data with short scan time. The DMI data were acquired with a standard spatial resolution of 15.6 μL (11×11×11 matrix over 27.5×27.5×27.5 mm³ FOV) but short scan times of 16 (NA = 4), 8 (NA = 2) and 4 (NA = 1) min. (A) $^2$H MR spectra from a tumor voxel, at NA = 4, 2 and 1; (B) MRI (tumor circled with the red dashed line; (C) concentration (Conc) and estimated SD maps given by spectral fitting (Fitted) and PRECISE-DMI (λ = 0.01), at corresponding number of averages in (A); (D) The concentration maps after applying anisotropic diffusion (AD), with gradient threshold of 10% and 20% (of the range of voxel values), to spectral fitting results, at NA = 1. Only the maps for Lac are shown.*

## **Discussion and Conclusion**

The proposed method PRECISE-DMI improved the visualization of the metabolic maps and the estimation precision of $^2$H-labeled metabolite concentrations, compared to the standard Fourier reconstruction, for DMI data in both simulation studies and in vivo experiments.

CNN-based SVE (PRECISE-DMI with λ = 0) achieved a ~1.7-fold improvement in SNR compared to the standard Fourier reconstruction in the simulation studies (Figure 5A). This improvement implies that the spatial resolution of DMI data can be increased (for example from 8 μL to 4.7 μL), or the minimal scan time can be shortened, without a penalty in estimation precision of the metabolite concentrations. Additionally, the estimated SD of CNN-based SVE approached the theoretical limit defined by the CRLB very closely (Figure 5B), which means that ultimate precision of unbiased estimation was achieved. Although the traditional model-based spectral fitting could also achieve this limit, using CNN-based SVE did not require optimization computation, and therefore avoided the nonconvergence problem caused by poor initialization and reduced the computation time significantly (from ~80s to ~1s in simulation studies). This improved speed will be important when processing much larger clinical datasets in the future. Furthermore, CNN-based SVE was shown to be able to handle significant $B_0$ and $B_1$ inhomogeneities (Figure 6). However, the noise-induced bias at poor SNR is regarded as a limitation.

In the simulation studies, a moderate incorporation of the edge-preserving regularization (PRECISE-DMI with $\lambda = 0.01$) helped to achieve a further improvement in SNR (~3-fold). This level of incorporation was able to provide metabolic maps with much better visual quality for in vivo DMI data acquired with high spatial resolution (Figure 9) or short scan time (Figure 10). Compared to other edge-preserving denoising techniques (e.g. anisotropic diffusion), PRECISE-DMI imposed edge prior from MRI instead of using automated edge-detection, therefore performed better when processing very low SNR DMI data.

However, the MRI-based edge-preserving regularization improved the estimation precision (SD) at the cost of introducing inaccuracies (bias). This problem was observed through the over-smoothing effect and most importantly the false negative tumor analyses in Figure 7 and Figure 9: a heavily weighted regularization ($\lambda = 0.04$) improved the precision significantly, but it made the tumor disappear because the image contrast of the tumor was not established in the MRI. The level of this inaccuracy depends on the tumor size and was especially strong when the tumor was only visible in one or a small number of voxels (Figure 8). On the other hand, a regularization that was weighted less ($\lambda = 0.004$) resulted in less inaccuracies but gave limited precision enhancement. A moderate level of regularization ($\lambda = 0.01$) gave a reasonable tradeoff between overall precision improvement and accuracy at the false negative tumor. The problem mentioned above is also a known problem for other techniques that used the edge-preserving regularization to process low SNR MRSI data (19,20,21). The fundamental problem is to find a good balance between the sensitivity enhancement and the potential to increase inaccuracies, which is controlled by the regularization parameter $\lambda$. Typically, $\lambda$ is tuned case-by-case based on the visual quality of the metabolic maps, making the techniques hard to be integrated into standard clinical tools that could be used for different patients. For this reason, future research will focus on finding a reliable level of spatial information incorporation (i.e. learning optimal values for $\lambda$) using a more advanced NN architecture, trained with paired MRSI and (multiparametric) MRI. Due to a limited amount of in vivo MRSI data, synthesizing MRSI data from publicly available multiparametric MRI dataset is currently under investigation (22).

Another limitation of this work is that PRECISE-DMI was only tested on an animal RG2 tumor model that showed very stark and relatively homogenous image contrast on MRI, making the edge localization more reliable than on MRI of human. On a human MRI, if the DMI is acquired during the course of therapy which involves resection and chemoradiation, appearance of the tumor will likely be much less homogenous. PRECISE-DMI will therefore need to be optimized (e.g. more advanced MRI preprocessing steps) to be compatible with clinical MRI data of brain tumors.

PRECISE-DMI can also be used for estimating concentrations of different compounds, e.g. acetate and choline, by modifying the frequencies and the relaxation time constants of resonances in the training FIDs. Additionally, in order to extend PRECISE-DMI to $^1$H MRSI, the training FIDs will be redesigned with additional challenges, such as including lipid and baseline signals, and introducing lineshape distortion terms to accommodate greater effects of $B_0$ inhomogeneity.

In summary, this work provides a rigorous evaluation pipeline for postprocessing techniques of DMI: evaluating improvement in SNR over existing methods, potential bias introduced by (spatial) prior information, effects of tumor size, closeness to the CRLB and performance in handling field inhomogeneities, etc. This pipeline quantifies the contributions/drawbacks of each of the prior information used in an algorithm and will function as a foundation for evaluating any future techniques.

## Acknowledgements

The research is supported by NIH grant R01EB025840, R01CA206180 and R01NS035193.